\documentstyle[aps,twoside,floats,epsf]{revtex}
\begin{document}
\draft
\draft 
\title{
Theory of Colossal Magnetoresistance in Doped Manganites
}
\author{ A.S. Alexandrov$^{1,*}$ and A.M. Bratkovsky$^{2,\dagger}$}

\address{$^{1}$Department of Physics, Loughborough University,
Loughborough LE11 3TU, UK\\
$^{2}$Hewlett-Packard Laboratories, 3500~Deer~Creek
Road, Palo~Alto, California 94304-1392 }
\date{October 22, 1998  }

\maketitle
\begin{abstract}
The exchange interaction of polaronic carriers with localized
spins leads to a
ferromagnetic/paramagnetic transition in doped
charge-transfer insulators  with strong electron-phonon coupling.
The relative strength of the exchange and electron-phonon interactions
determines whether the transition
is first or second order.
A giant drop in the number of current carriers during the transition,
which  is a consequence of local bound pair (bipolaron)
formation in the paramagnetic phase, is extremely sensitive to an external
magnetic field. Below the critical temperature of the transition,
$T_c$, the binding
of the polarons into immobile pairs competes with the ferromagnetic
exchange between polarons and the localized spins on Mn ions,
which tends to align the polaron moments and, therefore, breaks up
those pairs. The number of carriers abruptly increases below
$T_c$ leading to a sudden drop in resistivity.
We show that the carrier density collapse
describes   the colossal magnetoresistance  of  doped  manganites
close to the transition.
 Below $T_c$, transport occurs by polaronic
tunneling, whereas at high temperatures the transport is by
hopping processes. The transition is accompanied by a spike in
the specific heat, as experimentally observed.
The gap feature in tunneling spectroscopy is related to the bipolaron
binding energy, which depends   on the ion mass. This dependence
explains the giant isotope
effect of the magnetization and resistivity upon substitution of
$^{16}$O by $^{18}$O. It is shown also that the localization of polaronic
carriers by disorder {\em cannot} explain the observed huge sensitivity of
the transport properties to the magnetic field in doped manganites.

\end{abstract}

\section {Introduction}

The existence of a metal-insulator transition in lanthanum manganites
was established in the early 1950s\cite{van} and has been extensively
studied thereafter.
The transition is associated with unusual transport properties,
including large magnetoresistance in the vicinity of the
transition, studied in a family of doped manganites with perovskite
structure with the chemical formula Re$_{1-x}$D$_x$MnO$_3$, where Re
is the rare earth (Re = La, Pr, Nd), and D is the divalent metal (D =
Ca, Sr, Ba).
It is worth mentioning the early studies of the transition in
 La$_{1-x}$Pb$_x$MnO$_3$ \cite{searle}, followed by the studies of
Pr$_{1-x}$Ca$_x$MnO$_3$\cite{jirak},
Nd$_{0.5}$Pb$_{0.5}$MnO$_3$\cite{kusters},
La$_{0.67}$Ba$_{0.33}$MnO$_3$\cite{helmolt},
La$_{0.75}$Ca$_{0.25}$MnO$_3$\cite{chahara},
La$_{1-x}$Ca$_x$MnO$_3$\cite{jin,sch} (see review
\cite{ram}).
The recent resurgence of interest in these systems is related to the
demonstration of a very large negative magnetoresistance in thin
films\cite{helmolt,jin} [sometimes termed colossal magnetoresistance
(CMR)],  which immediately raised the possibility of technological
applications. The colossal magnetoresistance is not limited to doped
perovskite manganites, but was also observed in pyrochlore manganites,
chromium spinels \cite{ram}, and some other systems, like europium
compounds.

The metal-insulator transition in lanthanum
man\-ganites\cite{van,jin,sch}
has been traditionally attributed
to a `double exchange' mechanism, which results in
a varying band width of holes doped into the Mn$^{3+}$ $d$-shell
 as a function of the doping concentration and temperature\cite{dou}.
Recently it has been  realized \cite{mil}, however,  that the effective carrier-spin
exchange interaction
of the double-exchange model is too weak to lead to a  significant
reduction of
the electron bandwidth
and, therefore, cannot account for  the observed scattering rate
\cite{edw}
(see also Ref.~\cite{fisher}) or for localization induced
by slowly fluctuating spin configurations\cite{var}.
In view of this severe shortcoming of the double exchange model, it has
been suggested \cite{mil} that the essential physics of perovskite
manganites lies in the strong coupling of carriers to
the Jahn-Teller lattice distortion.
The argument \cite{mil} was that
in the high-temperature state the electron-phonon coupling
constant $\lambda$ is large
(so that the carriers are  polarons);
as temperature decreases the growing ferromagnetic order
increases the bandwidth and thus decreases $\lambda$
sufficiently for  metallic behavior to occur below
the Curie temperature $T_{c}$, in
accordance with polaron theory\cite{alemot}.
A giant isotope effect \cite{mul},
the sign anomaly of the Hall effect,
and the Arrhenius behavior of the drift  and Hall mobilities \cite{emi}
over a temperature range from $2T_{c}$ to $4T_{c}$ unambiguously
confirmed
the polaronic nature of the carriers in manganites.
Polaron hopping transport accounts satisfactorily for the  resistivity
in the paramagnetic phase\cite{emi}.

However, the known relation between magnetization and
transport below $T_c$ and the unusual magnetic ion dynamics
have prompted the conclusion that polaronic hopping is also the prevalent
conduction mechanism below $T_c$\cite{hundley}.
Low-temperature optical\cite{oki1,oki2,kim},
electron-energy-loss
(EELS)\cite{ju} and photoemission spectroscopies \cite{des}
showed that the idea \cite{mil,var}
of a `metalization' of manganites below $T_{c}$  is not
tenable.
A broad incoherent spectral feature \cite{oki1,oki2,kim,des} and a
pseudogap
in the excitation spectrum \cite{des,biswas,wei} were observed while
the coherent Drude weight
appeared to be {\em two orders} of magnitude smaller \cite{oki2} than
is expected for a metal, or even {\em zero} in the case of layered
manganites \cite{des}.
EELS \cite{ju} confirmed that  manganites
are charge-transfer  doped insulators
having $p$-holes as the current carriers rather than $d$ (Mn$^{3+}$)
electrons.
The photoemission and O $1s$ x-ray absorption spectroscopy of
La$_{1-x}$Sr$_x$MnO$_3$ showed  that the
itinerant holes doped into LaMnO$_3$ are indeed of oxygen $p$
character, and their coupling with the $d^4$ local moments on Mn$^{3+}$ ions
aligns the moments ferromagnetically\cite{saitoh}.
Moreover, measurements of the  mobility \cite{ram,htkim} do not
show any field dependence  and there are
significant deviations from Arrhenius behavior
close to $T_{c}$\cite{whi,emi}. The resistivity
calculated from the modified double-exchange theory
 is in poor agreement with the data  and the characteristic
theoretical
field ($\sim$15T) for CMR is too high
compared with the experimental one ($\sim$4T)\cite{mil}.
As a result, self-trapping  above $T_{c}$
and the idea of metalization below $T_{c}$
do not explain  CMR either.
Carriers retain their polaronic
character well below $T_{c}$, as  manifested also
in the measurements of resistivity and thermoelectric power
under pressure\cite{gud}.

Therefore, the experimental evidence
overwhelmingly suggests  that the low-temperature
phase of the doped manganites is not a metal, but a doped polaronic
semiconductor. The double exchange and the presence of polaronic
carriers
{\em are insufficient} to explain the physics of colossal
magnetoresistance. One can also add that there are known classes of
CMR materials where it is guaranteed that double exchange is
{\em non-existent}, like in pyrochlore manganites, chromium spinels
\cite{ram}, and other compounds.

In the present paper, we propose a new theory
of the ferromagnetic/paramagnetic phase transition accompanied by
a  current carrier density collapse (CCDC)  and  CMR.
Taking into account the tendency of polarons to form
local bound pairs (bipolarons)
as well as the  exchange
interaction of $p$ polaronic holes with $d$ electrons,  we find
a novel ferromagnetic transition driven by
non-degenerate polarons in doped
charge-transfer magnetic insulators. The crux of the matter is that
in the paramagnetic state above the critical temperature a large fraction
of polarons is bound into immobile pairs (bipolarons). As the
temperature decreases in the paramagnetic phase
($T>T_c$), so does the density of mobile polarons, and the resistivity
quickly increases with the decline of the number of carriers. With the
onset of ferromagnetic order at $T_c$, the situation changes
dramatically. As a result of the exchange interaction with the
localized Mn spins, the energy of one of the polaron spin sub-bands
sinks abruptly  below the energy of the bound pairs. The pairs break
up, the density of carriers (mobile polarons) jumps up, and the
resistivity suddenly declines, as observed experimentally. The
occurrence of the deep minimum
in  the carrier density close to the transition point, which we suggest
calling a {\em current carrier density collapse}, allows us
to explain the magnetization and
temperature/field dependence of the resistivity
of La$_{1-x}$Ca$_{x}$MnO$_{3}$  close to $T_{c}$ as well as the giant
isotope effect, the unusual tunneling gap, and the specific heat anomaly.

\section {Ferromagnetic transition in doped
manganites}

The Hamiltonian containing the physics compatible
with the experimental observations mentioned above is
\begin{eqnarray}
{\cal H}&=&\sum_{k,s} E_{ {\bf k}} h^{\dagger}_{{\bf k}s} h_{{\bf k}s}
 - \frac{J_{pd}}{2N} \sum_{{\bf k},j}m_{\bf k}S^z_j + {\cal
H}_{\rm sf}
 +{\cal H}_{\rm Hund}\nonumber\\
& + &(2N)^{-1/2}\sum_{{\bf k,q},s}\hbar\omega_{\bf q}
\gamma_{\bf q} h^{\dagger}_{{\bf k+q}s} h_{{\bf k}s}(b_{\bf
q}  -  b^\dagger_{\bf-q})
\nonumber\\
&+&\sum_{\bf q} \hbar\omega_{\bf q}
(b^\dagger_{\bf q}b_{\bf q} + 1/2),
%
\label{eq:ham}
\end{eqnarray}
where $E_{ {\bf k}}$ is  the local density approximation({\sc LDA})
energy dispersion\cite{pic},
$ h_{ {\bf k}s}$ is
the annihilation hole operator
 of a (degenerate) $p$ oxygen band with spins
$s=\uparrow$ and $\downarrow$,
$J_{pd}$ is the exchange
interaction of $p$ holes with four $d$ electrons of
the Mn$^{3+}$ ion at the site $j$, $m_{{\bf k}}\equiv
h^{\dagger}_{ {\bf k}\uparrow} h_{{\bf k}\uparrow}-
h^{\dagger}_{ {\bf k}\downarrow} h_{{\bf k}\downarrow}$,
$S^z_j$ is the $z$-component of Mn$^{3+}$ spin, which is
$S=2$ due to the
strong Hund coupling, ${\cal H}_{\rm Hund}$,
of the four $d$-electrons on Mn$^{3+}$ sites,
$N$ is the number of unit cells.
The two last terms of the Hamiltonian describe the
coupling of $p$ holes  with phonons and the phonon energy,
respectively ($\gamma_{\bf q} = -\gamma^*_{\bf -q}$ is the coupling
constant \cite{alemot}).
The Hamiltonian
also contains spin-flip processes, ${\cal H}_{\rm sf}$, like
$S_j^{+}h^\dagger_{\bf k'\downarrow}h_{\bf k\uparrow}+h.c.$,
and terms with non-diagonal components of
the polaron magnetization operator $m_{\bf k'k} =
h^\dagger_{\bf k'\uparrow}h_{\bf k\uparrow}
-  h^\dagger_{\bf k'\downarrow}h_{\bf k\downarrow}$, which are not
essential for our discussion.
 If the holes
were doped into the $d$ shell instead of the $p$ shell, the Hamiltonian
would be similar to the Holstein $t-J$ model \cite{fes}
with about the same physics of CMR as proposed below.


The essential results are readily obtained
within the Hartree\--Fock approach
for the exchange interaction \cite{yosida} and the Lang-Firsov polaron
trans\-formation\cite{lfirsov}
which removes terms of first order in the electron-phonon
interaction in Eq.~(\ref{eq:ham}),
$\tilde{H}=e^U H e^{-U}$, where
\begin{equation}
U = \sum_{j{\bf q}s}h^\dagger_{js} h_{js} u_{j\bf q}
(b^\dagger_{-{\bf q}} + b_{\bf q}),
\end{equation}
$ h_{js}$ = $N^{-1/2}\sum_{\bf k}h_{{\bf k}s} \exp(\imath {\bf k\cdot
R}_j)$, $ u_{j\bf q}$ = $(2N)^{-1/2} \gamma_{\bf q}\exp(\imath {\bf
q\cdot R}_j)$, and ${\bf R}_j$ is the lattice vector.

With the use of this transformation one finds spin-polarized $p$ bands
\begin{equation}
\epsilon_{ \bf k\uparrow(\downarrow)}
=\epsilon_{\bf k}  -(+) \frac{1}{2}J_{pd} S \sigma -(+)\mu_{B}H.
\label{eq:epol}
\end{equation}
Here
\begin{equation}
\epsilon_{\bf k} = \frac{1}{N}\sum_{i,j}t_{ij}
e^{\imath {\bf k}\cdot({\bf R}_i-{\bf R}_j)}
e^{-g^2_{ij}} \approx E_{\bf k}e^{-g^2}
\label{eq:edef}
\end{equation}
where
\begin{equation}
g^2_{ij} = \frac{1}{2N}\sum_{\bf q}|\gamma_{\bf q}|^2
\left[ 1-\cos{\bf q}\cdot({\bf R}_i - {\bf R}_j)\right]
\coth\left({\hbar\omega_{\bf q}\over{2k_BT}}\right)
\label{eq:g2}
\end{equation}
where $g^2 \sim \gamma^2$ is the characteristic value of $g^2_{ij}$
and $\omega_{\bf q}$ is the
phonon frequency.
 Equation (\ref{eq:g2})
describes the polaronic band narrowing
\cite{lfirsov} and the isotope effect\cite{mul}  .
The bare hopping integrals $t_{ij}$
define the unrenormalized {\sc LDA} (local density approximation) band
dispersion in the initial Hamiltonian (\ref{eq:ham})
$E_{\bf k} = \frac{1}{N}\sum_{i,j}t_{ij}
\exp[\imath {\bf k}\cdot({\bf R}_i - {\bf R}_j)]$.
In Eq.~(\ref{eq:epol}) $\sigma$ is the
normalized thermal average of the Mn spin,
found from the equations below;
$H$ is the external magnetic field, and $\mu_{B}$ is the Bohr magneton.
The $p-d$ exchange interaction depends only on total (average)
magnetization because we
assume that the system is homogeneous. In addition to band narrowing,
the bands shift rigidly down, the value of this {\em polaron shift}
$E_p$ is
\begin{equation}
E_p = \frac{1}{2N} \sum_{\bf q} \hbar\omega_{\bf q}|\gamma_{\bf q}|^2.
\end{equation}

The ions  Mn$^{3+}$ are subject
to a molecular field
$J_{pd}m/(2g_{\rm Mn}\mu_{B})$, according to (\ref{eq:ham}),
and their magnetization  $\sigma\equiv  \langle S^z_{\bf n}\rangle/S$
is given by
\begin{equation}
\sigma=B_{S}\left({J_{pd}m + 2g_{\rm Mn}\mu_{B} H \over{2 k_B
T}}\right),
\end{equation}
with $m$ the magnetization of holes determined as
\begin{equation}
m \equiv  {1\over N} \sum_{\bf k}\langle m_{\bf k}\rangle
=\int d\epsilon N^{\rm (p)}(\epsilon)
\left[ f_{\rm p}(\epsilon_{\bf k\uparrow})
-f_{\rm p}(\epsilon_{\bf k\downarrow})\right].
\label{eq:mint}
\end{equation}
Here $B_{S}(x) =\- [1+1/(2S)]\- \coth[(S+1/2)x]\- -[1/(2S)]\coth(x/2)$
is the Brillouin function, $g_{\rm Mn}$ the Lande $g$-factor for
Mn$^{3+}$ in a manganite,
$N^{\rm(p)}(\epsilon)$  the density of states in the narrow polaron band,
and $f_{\rm p}(\epsilon_{{\bf k}s})
=[y^{-1}\exp(\epsilon_{{\bf k}s}/k_{B}T)+1]^{-1}$
the Fermi-Dirac distribution function
with $y=\exp(\mu/k_B T)$ determined by the chemical potential $\mu$.
Note that for $J_{pd}<0$ (antiferromagnetic coupling) the main system
of equations (\ref{eq:n})-(\ref{eq:sig}) remains the same after a
substitution $J_{pd} \rightarrow |J_{pd}|$.

Along with the band narrowing effect,
the strong electron-phonon interaction binds two polarons into
a local pair (bipolaron), as described in detail in Ref.~\cite{alemot}.
These bipolarons are practically
immobile in manganites because of
the strong electron-phonon interaction,
in contrast with cuprates, where
 bipolarons are mobile  and  responsible
for in-plane transport\cite{alebra}, owing to their geometry \cite{ale}
and their moderate coupling with phonons\cite{chak}.

If these bound pairs are extremely local objects, i.e. two holes on the same
oxygen, then they will form a singlet. If, however, these holes are
localized on different oxygens, then they may well have parallel spins
and form a {\em triplet} state. The latter is separated from the
singlet state by some exchange energy $J_{st}$, with some interesting
consequences discussed below.
Because of their zero spin, the only role of the
singlet bipolarons in manganites
is to determine the chemical potential $\mu$, which can be found with
the use of the
total doping density per cell $x$\cite{alebra}.

%
%
The interplay between the
localization of $p$-holes into  bipolaron pairs
and the exchange interaction
with the Mn $d^4$ local moments
is responsible for CMR.
The density of these pairs has a sharp
peak at the ferromagnetic transition when the system is cooled down through
the critical temperature $T_c$.
As the system is cooled, but is still in a paramagnetic state above
$T_c$, an increasing fraction of the polarons forms immobile pairs
(bipolarons), and the resistivity of the system increases.
Below $T_c$, the binding
of polarons into immobile pairs competes with the ferromagnetic exchange,
which tends to align the polaron moments and, therefore, breaks
those pairs apart. The number of carriers abruptly increases below
$T_c$ leading to a sudden drop in resistivity.
These competing interactions lead to the unusual
behavior of CMR materials and the extreme sensitivity of their transport to
external fields.

To prove the point, we shall find the thermodynamic potential and
solve for its extremal value to find the equation of state for the polarons.
The thermodynamic  potential
$\Omega$
\begin{equation}
\Omega = \Omega_{\rm p} + \Omega_{\rm bp} + \Omega_{S}
+ \frac{1}{2}J_{pd}S\sigma m,
\label{eq:Om}
\end{equation}
has contributions from polarons, bipolarons, localized Mn$^{3+}$
spins, and the double-counting term, respectively.
For the polarons
\begin{equation}
\Omega_{\rm p} = -k_BT\sum_s \int d\epsilon N_s^{\rm (p)}(\epsilon)
\ln\left(1+ye^{-\epsilon/k_BT}\right),
\label{eq:Ompol0}
\end{equation}
where $N_s^{\rm (p)}(\epsilon)$ is the density of spin polarized states in
the polaron band, and $y\equiv\exp(\mu/k_BT)$.
We can easily estimate the integral,
assuming that the critical temperature of the ferromagnetic transition
is comparable with the polaron and bipolaron bandwidth\cite{twfac}. Then
(bi)polarons are not degenerate in the relevant temperature range,
$f_{\rm p}\simeq y\exp(-\epsilon /k_{B}T)$ and
$f_{\rm bp}\simeq y^2 \exp[(\Delta - \epsilon )/k_{B}T]$, and we get
\begin{equation}
\Omega_{\rm p} = -2y\nu k_BT\cosh{J_{pd}S\sigma+g\mu_B H \over{2k_BT}},
\label{eq:Ompol1}
\end{equation}
where $\nu(=3)$ is the degeneracy of the polaron $p$ band.

Polarons, bound in bipolarons with a binding energy $\Delta$, give a
contribution
\begin{equation}
\Omega_{\rm bp} = -k_BT \ln\left( 1 + \nu^2 y^2 De^{\Delta/k_BT}\right),
\label{eq:Ombp}
\end{equation}
where $D$ accounts for the presence of triplet bipolarons (see below
Sec.\ref{sec:triplet}). We shall consider here a simple case when the
separation of the triplets from the singlets, $J_{st}$, is much larger
than the critical temperature. In this case $D=1$.

Finally, for the localized spin contribution we will have
\begin{equation}
\Omega_{S} = -k_BT \ln{ \sinh(S+\frac{1}{2})\eta
\over{\sinh\frac{1}{2}\eta}},
\end{equation}
with $\eta = (\frac{1}{2}J_{pd}m +  g_{\rm Mn}\mu_B H )/k_BT$.

The density of polarons $n=-(\partial \Omega_{\rm p}/\partial \mu)_T$
is found from the condition that the total number
of carriers is given by the doping concentration $x$\cite{alebra}:
\begin{equation}
x= -(\partial \Omega/\partial \mu)_T,
\label{eq:x}
\end{equation}
whereas one can find equations for the magnetization and the
normalized spin $\sigma$ from  the following conditions:
\begin{equation}
(\partial \Omega/\partial \sigma)_T=
 (\partial \Omega/\partial m)_T=0.
\label{eq:eqlbr}
\end{equation}
Thus, we obtain the following main system of mean field equations, assuming
for a moment that the contribution from triplet bipolarons is small ($D=1$):
%
%
\begin{eqnarray}
n&=& 2\nu y  \cosh[(\sigma+h)/ t],\label{eq:n}\\
m&=&n \tanh[(\sigma+h)/ t],\label{eq:m}\\
\sigma&=&B_2 [(m+4h)/(2 t)],\label{eq:sig}
\end{eqnarray}
and
\begin{equation}
y^{2}= {x-n\over{2\nu^2}} \exp(-2\delta/t),
\label{eq:y}
\end{equation}
which follows from (\ref{eq:x}). When triplet bipolarons become
important, one should replace Eq.~(\ref{eq:y}) by the more accurate
Eq.~(\ref{eq:yt}).
Here we use the dimensionless temperature $t=2k_{B}T/(J_{pd}S)$,
magnetic field $h=2\mu_{B}H/(J_{pd}S)$, the bipolaron
binding energy $\delta\equiv \Delta/(J_{pd}S)$, and
$\nu(=3)$ is defined after Eq.~(\ref{eq:Ompol1}).

The polaron density $n$ is determined by Eqs.~(\ref{eq:n}),
(\ref{eq:y}) with $\sigma=0$ above
$T_{c}$. At the critical temperature,
the polaron density has a minimal value
$n_{c}\simeq (2x)^{1/2}\exp(-\delta/t_{c})$, it then grows
with temperature and saturates
at $n=(1+2x)^{1/2}-1$ at large temperatures.
This is reminiscent of ordinary semiconductor behavior.

\section{Competing interactions and carrier density collapse}

\begin{figure}[t]
\epsfxsize=4in
\epsffile{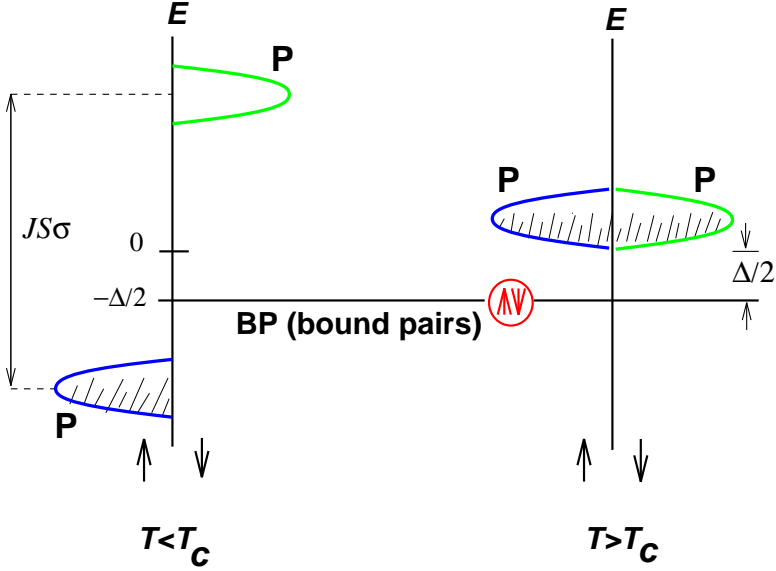}
\vspace{.2in}
\caption{
Schematic of free polaron (P) and polaron bound pair (BP) densities of
states at temperatures below and above $T_c$ for up ($\uparrow$) and
down ($\downarrow$) spin moments. The pairs (BP) break below $T_c$
if the exchange $J_{pd}S$ between $p$-hole polarons and Mn $d^4$ local spins
exceeds the pair binding energy $\Delta$, as in the case shown.
The exchange interaction of polarons with the localized
spins sets in below $T_c$,
the spin-up polaron sub-band sinks abruptly below the
bipolaron band, causing the break-up of the immobile bipolarons (left panel).
A sudden drop (collapse) of the density of the current carriers
(polarons) in the vicinity of the ferromagnetic transition is the
cause of a large peak in resistivity and colossal magnetoresistance.
\label{fig:BndDgr}
}
\end{figure}

The remarkable  observation
is that there is a sharp increase of the polaron density (and the
conductivity) at  temperatures below $T_{c}$.
The polaron density approaches the total density $x$
at $T\rightarrow 0$ if $\delta\equiv \Delta/J_{pd}S <1$,
as one can see from Eq.~(\ref{eq:n}) with a saturated magnetization $\sigma=1$.
The physical origin of the unusual minimum of
the current carrier density at $T_{c}$ lies
in the instability of bipolarons below $T_{c}$ due to the exchange
interaction of polarons with Mn $d$ electrons. The spin-polarized polaron
band falls below the bipolaron band, so that all
carriers are unpaired at $T=0$ if $J_{pd}S\geq \Delta$.
The evolution of the Hartree-Fock bands with temperature, which
corresponds to this behavior, is illustrated in Fig.~\ref{fig:BndDgr}.
Note that at all $T>T_c$ the position of the polaron bands is
fixed at $\Delta/2$ above the
bipolaron band, since there $\sigma=m=0$ (\ref{eq:epol}). Their
population depends on temperature via the chemical potential.
The exchange interaction of polarons with the localized spins sets in
at $T_c$, and in the low-temperature ferromagnetic phase
one of the polaron spin sub-bands sinks abruptly below the
bipolaron band, causing the break-up of the immobile bipolarons.
This interesting feedback mechanism can result in either a continuous or
discontinuous ferromagnetic transition, as follows from a
simple analysis below.

Linearizing Eqs.~(\ref{eq:n})-(\ref{eq:sig})
 with respect to $\sigma$ and $m$  near $T_{c}$,
we find the critical temperature in zero magnetic field
\begin{equation}
t_{c}= (n_{c}/2)^{1/2},
\end{equation}
where the  polaron density at the transition $n_{c}$ is determined by
\begin{equation}
n_c^{1/2} \ln{2(x-n_c)\over{n_c^2}}=2^{3/2}\delta.
\label{eq:nc}
\end{equation}
It is easy to see that this transcendental equation has  solutions
only for $\delta$ below some critical value
$\delta_{c}(x)$. This means that  for $\delta>\delta_{c}(x)$
the ferromagnetic phase transition is first order
with jumps of the polaron density and the magnetization\cite{ab_cmr1},
as has been observed\cite{kuw}.
The transition is continuous when $\delta<\delta_{c}(x)$.
The numerical solution of the system Eqs.~(\ref{eq:n})-(\ref{eq:y})
defines the crossover between first- and second-order phase
transitions \cite{ab_cmr1}.

\begin{figure}[t]
\epsfxsize=4in
\epsffile{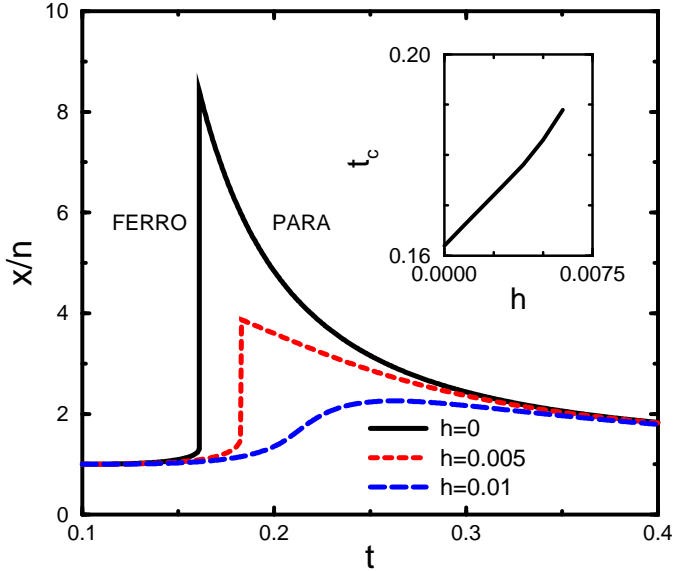 }
\caption{Inverse polaron density $x/n$
in a doped charge-transfer insulator
for different magnetic fields $h\equiv g\mu_B H/J_{pd}S$,
$\Delta/J_{pd}S=0.5$, doping $x=0.25$. $\Delta$ is the pair binding
energy,
$J_{pd}S$ is the exchange energy of the O~$p$ hole polarons
with Mn~$d$ localized spins. For other notations see text.
Note that the transition is a strong first order, and then becomes
continuous
when the external magnetic field exceeds some critical
value. Inset: temperature of the phase transition as a function
of external magnetic field.
\label{fig:1on}
}
\end{figure}

A relatively weak magnetic field has a drastic effect on
the inverse carrier density, $1/n$, near the
first order phase transition, or second order phase transition close to
first order.
A field equal to only $0.005J_{pd}S/(2\mu_B)$ reduces the carrier density
collapse by  more than a factor of two,
and a field of $0.01J_{pd}S/(2\mu_B)$ changes the transition into the
continuous one, Fig.~\ref{fig:1on}.
This behavior directly relates to the colossal magnetoresistance found
in doped manganites, as we shall discuss in the following section.

One can draw an analogy of this situation with singlet magnetism,
e.g. in Pr compounds \cite{singmag}. In this case the
ground state of magnetic ions is singlet. Depending on the ratio
between the exchange constant and singlet-triplet(doublet) energy gap
produced by crystal-field splitting,
there exist first- or second-order phase transitions into a
ferromagnetic state. In our case the triplet states become important
when $J_{st} \lesssim \Delta$, the larger statistical weight of
triplet bipolarons
leads to a deeper minimum in the density of polarons at the critical
temperature, and, therefore, to a larger jump in resistivity. The
effect of polaron binding in a triplet state will be discussed below.

\section{Colossal magnetoresistance }

\begin{figure}[t]
\epsfxsize=4.6in
\epsffile{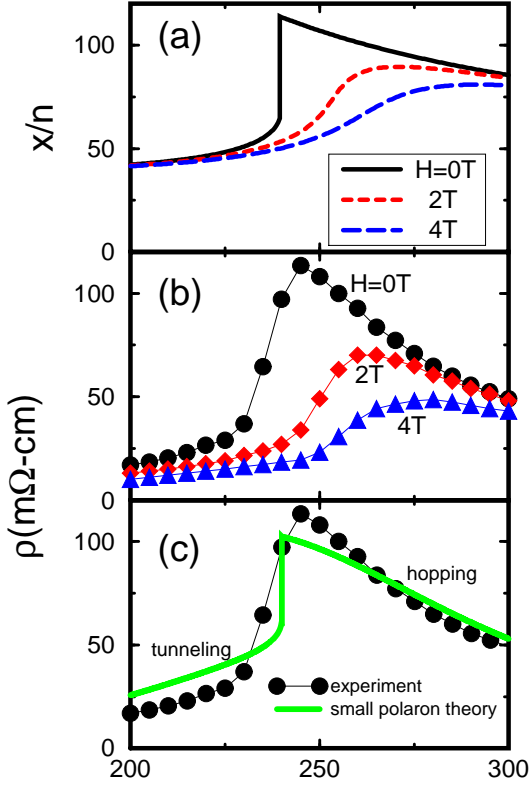 }
\caption{
Resistivity of La$_{0.75}$Ca$_{0.25}$MnO$_{3}$
calculated within the present theory
for $\Delta=900$K, $J_{pd}S=2250$K for a temperature independent mobility
(a). The experimental results [8] are shown
on panel (b).  Note the extreme sensitivity
of the theoretical resistivity to the external magnetic field (a),
also observed experimentally for the doped manganite (b) (thin solid
line is a guide to the eye).
Panel (c): Resistivity calculated with a temperature dependent
mobility according to Eq.~(26) with $\omega_0=50$meV and
$E_a=300$meV and temperature dependent
polaron density from panel (a) compared with the experimental
results. Note the
crossover of the transport mechanism
from low-temperature tunneling to high-temperature hopping
at about the transition temperature. For notations see caption
to Fig.~\ref{fig:1on}.
\label{fig:rte}
}
\end{figure}

As a result of the carrier density collapse, the resistivity
$\rho=1/(en \mu_{\rm p})$
has a sharp maximum, which is
extremely sensitive to the magnetic field in the vicinity of $T_{c}$.
In fact, our theory, Eqs.~(\ref{eq:n})-(\ref{eq:y}),
describes all the major features of the
temperature/field dependence of $\rho(T)$\cite{sch}, with
a temperature and the field independent polaron drift mobility
$\mu_{\rm p}$
in the experimental range of the magnetic field,
Fig.~\ref{fig:rte}(a),(b).
It gives the correct magnitude of the effect on resistivity and explains
the extreme sensitivity to external magnetic fields.
This suggests that {\em current carrier density collapse} is the
origin of CMR.
%
%

In general, one has to take into account
the temperature dependence of the
polaron mobility to extend our theory for temperatures far away from
the transition.
At high temperatures, the mobility $\mu_{\rm p}$ of polarons is dominated by
hopping events since the polaron narrowing factor $g^2$ grows linearly
with $T$, making tunneling in a narrow polaron band virtually impossible at
$k_BT>\hbar\omega_0/2$, where $\omega_0$ is the characteristic phonon
frequency\cite{lfirsov}.
A simple estimate for the so-called {adiabatic} hopping conductivity
together with the Einstein relation between diffusion constant and
mobility immediately yields
\begin{equation}
\mu_{\rm p}^{\rm (hop)} \sim {\mu_0\over {2 \pi}}
{\hbar \omega_0\over{k_BT}}\exp(-E_a/k_BT),
\label{eq:hop}
\end{equation}
where $\mu_0=ea^2/\hbar$ is the characteristic mobility
(one can estimate $a$ as the O-O distance in manganites),
 and $E_a$ is the activation energy
for the hopping.
Tunneling mobility is given by
\begin{equation}
\mu^{\rm (tun)}_{\rm p} = \mu_0 {{\bar t}^2e^{-2g^2}\over{\hbar k_BT}}\tau,
\label{eq:mutun}
\end{equation}
with the relaxation time $\tau$ estimated by Lang and Firsov\cite{lfirsov}
\begin{equation}
\tau \approx (E_a/\bar t)^4 [\Delta \omega/\omega_0^2]
\sinh^2(\hbar\omega_0/2k_BT),
\end{equation}
where $\bar t$ is the characteristic bare hopping integral $t_{ij}$,
and $\Delta\omega$ is the phonon dispersion.
The resistivity is then given by
\begin{equation}
\rho = 1/\sigma, \hspace{.1in} \sigma = ne(\mu^{\rm (tun)}
+\mu^{\rm (hop)}).
\label{eq:rho}
\end{equation}
With our low polaron density at the transition
 the polaron mobility is
$\mu_{\rm p}=0.2$~cm$^2$/Vs
for $x=0.33$ \cite{emi}, and about $0.03$~cm$^2$/Vs for $x=0.25$\cite{sch} that
lies in the range typical of polaronic conductors like TiO$_2$ at room
temperature \cite{alemot}.
We have fitted the observed resistivity to the above expression
[Fig.~\ref{fig:rte}(c)]
using for $\omega_0$ a value of 50 meV, which is close to  the phonon
cutoff in LCMO (50-70 meV\cite{aeppli}).

The fit indicates that the activation energy
is close to $E_a=300$ meV.
A crossover from tunneling to hopping occurs around the critical
temperature $T_c$, which is not very different from
 $\hbar\omega_0/2k_B$\cite{peterzhao}.
Agreement with the experiment (Fig.~\ref{fig:rte}(c))
supports the idea that the temperature dependence
of the resistivity is due primarily
to CCDC. The temperature dependence of the small polaron mobility then
allows the resistivity far away from the transition $both$
above and below T$_{c}$ to be explained.

\section{ Anomalous specific heat }
\begin{figure}[\!h]
\epsfxsize=4in
\epsffile{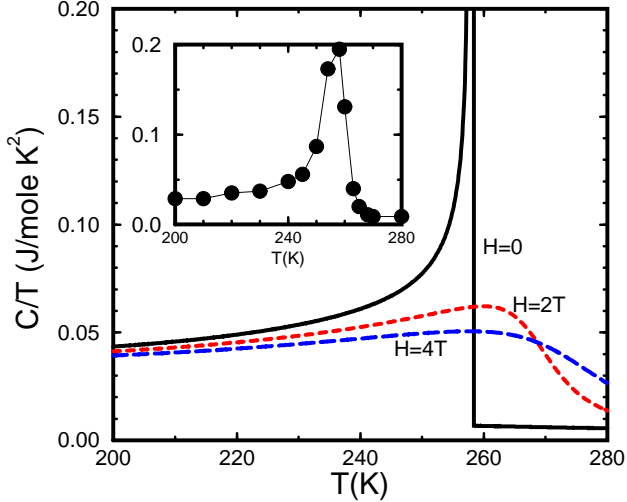}
\caption{
Calculated anomalous part of the specific heat for different values of
the magnetic field $H$. Inset:
experimental results for La$_{0.67}$Ca$_{0.33}$MnO$_3$
[42] (thin solid line is a guide to the eye).
\label{fig:cp}
}
\end{figure}

The carrier density collapse is also evident through anomalies in
thermodynamic quantities. Indeed, we have shown above that the
ferromagnetic
transition is first order, or second order close to first order,
as observed. The thermodynamic potential changes rather
abruptly in the vicinity of the phase transition and this results in a
sharp peak in the specific heat $C$, Fig.~\ref{fig:cp}, which has been
observed\cite{heatcap}. Note that this is {\em not} a result of critical
fluctuations as suggested earlier\cite{heatcap},
since they are absent or severely suppressed when the phase transition is
first order, or close to it.
We see that our theory is in quantitative agreement with the
experiment for this anomalous thermodynamic quantity.

\section{Triplet bipolarons}
\label{sec:triplet}

Let us now discuss the modification which arises if we include exchange between
O-holes bound into bipolarons. This exchange generally induces a
splitting $J_{st}$
between singlet and triplet states of the bipolaron. This changes
somewhat the thermodynamic potential of the bipolarons,
since the triplet is subject to a Zeeman splitting.
The factor $D$ is then
\begin{equation}
D = 1 + e^{-J_{st}/k_BT} \sinh(3\xi/2)/\sinh(\xi/2)
\label{eq:D}
\end{equation}
as it accounts for thermal excitations of singlet bipolarons into the triplet
state, separated from the singlet by the energy $J_{st}$. The parameter
\begin{equation}
\xi = (\tilde J_{pd} S \sigma + V_{bp} m
+  g\mu_B H)/k_BT,
\label{eq:xi}
\end{equation}
depends on exchange interaction of the bipolarons with Mn$^{3+}$ spins
given by the exchange constant $\tilde J_{pd}$, and delocalized
polarons, given by the exchange constant $V_{bp}$.
Note that
$D=4$ at $J_{st}/k_BT \ll 1$, whereas $D=1$ for $J_{st}/k_BT\gg 1$,
which reflects the higher statistical weight of triplet states compared
to singlets.

\begin{figure}[t]
\epsfxsize=4in
\epsffile{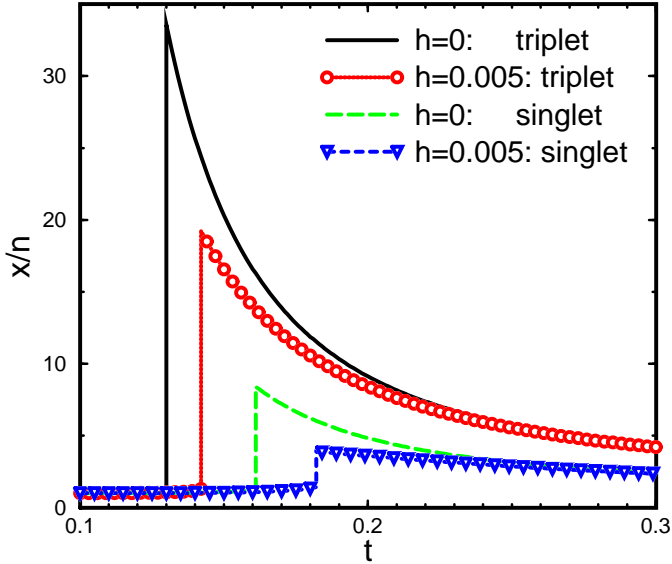}
\caption{Inverse polaron density $x/n$ for different magnetic fields
for a system with triplet and singlet bipolarons
versus temperature $t\equiv 2k_BT/J_{pd}S$.
($\Delta/J_{pd}S=0.5$, doping $x=0.25$, and we assume $J_{st}\ll \Delta$).
The jump in carrier density is much larger in a system with triplet
bipolarons, but the critical temperature and sensitivity of the
critical temperature to the magnetic field is lower in comparison with
singlet bipolarons. For notations see caption
to Fig.~\ref{fig:1on}.
\label{fig:ST}
}
\end{figure}

It is assumed, as is usually the case, that the triplet states
lie higher in energy than the singlet state, $J_{st}>0$. If the
singlet-triplet splitting becomes smaller than the gap, $J_{st} \lesssim
\Delta$, then, because of a higher number of the triplet states, their
thermal population leads to a deeper minimum in the density of polarons
and, therefore,
to a larger jump in resistivity (Fig.~\ref{fig:ST}). 
The dependence of
the population of the triplet states on external field makes the
system somewhat less sensitive to the field.
We make an essential assumption that $J_{st} >0$ and that the exchange
between spins on Mn and triplet bipolarons, $\tilde J_{pd}$, is
suppressed to values $\ll J_{pd}$ because the bipolarons are strongly
localized [we also expect that the exchange constant $V_{bp}$  is
the smallest one in  (\ref{eq:xi})].
Otherwise, the triplet bound pairs, if they were formed in the
paramagnetic phase, can survive in the ferromagnetic phase thus
reducing or eliminating the carrier density collapse.

The equation (\ref{eq:y}) is  changed to read
\begin{equation}
y^{2}= {x-n\over{(2-x+n)\nu^2 D}} \exp(-2\delta/t),
\label{eq:yt}
\end{equation}
whereas the main system of equations (\ref{eq:n})-(\ref{eq:y}) remains the same.
$D$ is given by Eq.~(\ref{eq:D}).
The effect
of triplet bipolarons on the thermodynamics of doped manganites becomes
insignificant when $J_{st}/\Delta > 1$,
and the results are similar to the case
when only singlet bipolarons are involved.
The  polaron density at the transition $n_{c}$ is determined by
\begin{equation}
n_c^{1/2} \ln{2(x-n_c)\over{Dn_c^2[1-(x-n_c)/2]}}=2^{3/2}\delta,
\end{equation}
which is similar to the case of singlet polarons and also indicates a
crossover from first- to second-order phase transition.
We compare the carrier density collapse in a system with triplet
bipolarons to that with singlet bipolarons alone in Fig.~\ref{fig:ST}.
The jump in the carrier density at the transition is a few times larger
in this case as compared to singlet bipolarons. At the same time
the critical temperature shifts to lower values, and the sensitivity to
external magnetic field slightly reduces.

\section{Tunneling gap and giant isotope effect}

 Recent tunneling measurements have shown that in the vicinity of $T_c$
a gap in the quasiparticle spectrum opens up\cite{biswas,wei}.
Again, it is difficult to reconcile this gap with the notion
of a (half-)metallic ferromagnetic state below $T_c$\cite{amb}.
In half-metallic ferromagnets, like CrO$_2$ or Fe$_3$O$_4$, there is a
band gap for electron states of only {\em one} spin direction. The
opposite spin
electrons have no gap  at the Fermi level,
similar to a standard
metal situation. These states will contribute to tunnel current as in
conventional metals, so that there would be {\em no} such
temperature dependent gap
feature in the tunnel spectroscopy\cite{amb} like the one observed for the
doped manganites\cite{biswas}.

\begin{figure}[t]
\epsfxsize=4in
\epsffile{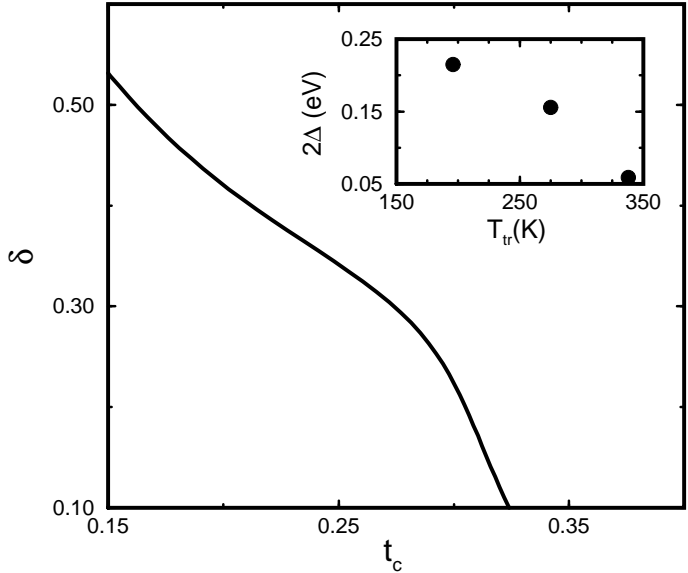 }
\caption{Relation between the gap $\delta\equiv \Delta/J_{pd}S$
and the critical temperature
$t_c\equiv 2k_BT_c/J_{pd}S$ calculated from the present theory.
Inset: tunneling gap
in the density of states for samples with different temperatures
of the transition: La$_{0.8}$Ca$_{0.2}$MnO$_3$, $T_{tr}$=196K;
(NdLa)$_{0.73}$Pb$_{0.27}$MnO$_3$ ($T_{tr}$=275K);
La$_{0.7}$Pb$_{0.3}$MnO$_3$ ($T_{tr}$=338K) [24].
For notations see caption to Fig.~\ref{fig:1on}.
\label{fig:gap}
}
\end{figure}

We note that within the  framework of our theory there should be a
temperature dependent gap $\Delta$ related to the breakdown of a
bipolaron
into two polaronic carriers. The density of bipolarons
peaks at $T_c$, whereas the polaron density dips there (Figs.~1, 2) and,
therefore, the gap feature in the tunneling $I-V$ curves will be most
pronounced in this region, as observed\cite{biswas}.
Spin-polarized polarons will provide a gapless background for
tunneling current, which is least important in the vicinity of the
transition temperature. We note that {\sc STM} should also be sensitive
to the presence of the one-particle charge-transfer gap between filled
Mn $d$ and empty O $p$ states.
In addition to the temperature dependence, we
can predict how the gap feature will depend on the critical temperature
of the transition (Fig.~\ref{fig:gap}). 
%
%
Namely, as already follows
from our discussion,
with the increase of $\Delta$ the critical temperature $T_c$
goes down\cite{ab_cmr1}.
Very similar behavior has indeed been observed experimentally on
samples with different critical temperatures
(Fig.~\ref{fig:gap}, inset)\cite{biswas}.

\begin{figure}[t]
\epsfxsize=4in
\epsffile{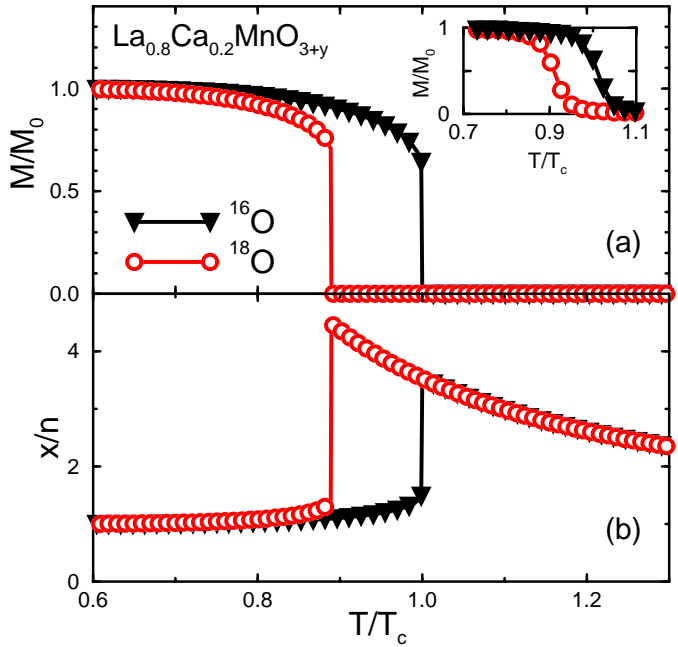 }
\caption{
Isotope effect on magnetization (a) and inverse carrier density (b)
of La$_{0.8}$Ca$_{0.2}$MnO$_{3+y}$
calculated in the present theory.
Inset: experimental results [16].
Substitution $^{16}$O$\rightarrow^{18}$O leads to increased
resistivity (b).
\label{fig:Iso}
}
\end{figure}

The giant isotope effect
in La$_{0.8}$Ca$_{0.2}$MnO$_3$, where a shift of -21K in $T_c$ was
observed
as a result of  $^{16}$O to $^{18}$O substitution\cite{mul}, is
quantitatively explained within our approach. Namely, the gap is given
by
\cite{alekab}
\begin{equation}
\Delta=
2E_p - V_C - \frac{1}{2}W,
\label{eq:del}
\end{equation}
where $E_p$ is the polaron level shift, $V_C$ is
the Coulomb repulsion between bound polarons, and $W =  W_0 \exp(-g^2)$
is the polaron bandwidth renormalized from the bare value $W_0$
with the electron-phonon interaction constant $g^2$\cite{alemot}.
The {\em only} quantity in (\ref{eq:del}) that depends on ionic mass is
the polaronic exponent $g^2 = \gamma E_p/(\hbar \omega)\propto
M^{1/2}$\cite{alemot}, where $\gamma <1$ is a numerical coefficient
depending on the radius of the electron-phonon interaction\cite{ale}.
As immediately follows from this relation, isotope substitution will
change
the gap $\Delta$ in the following way
\begin{equation}
\Delta_{18} = \Delta_{16} + Wg_{16}^2(\sqrt{18/16} -1),
\label{eq:iso}
\end{equation}
where indices mark the quantities for the corresponding isotopes of
oxygen.
According to (\ref{eq:iso}) $\Delta_{18}$
is {\em always} larger than $\Delta_{16}$.
This automatically leads to a lowering of $T_c$ as a result of the
isotope substitution, as observed\cite{mul}, in Fig.~\ref{fig:Iso}.
%
%
The resistivity, on the other hand,
is {\em larger} in $^{18}$O substituted samples, and this correlation
seems to be supported by recent experiments\cite{zhougud}.
Note that the {\em single parameter} defining the isotope effect on the
magnetic transition and the resistivity jump is $Wg_{16}^2$, since
neither $E_p$ nor $V_C$ depend on the ion mass.

\section{Localization of polarons by disorder}

We have also studied the localization of $p$-holes
due to a random field with a
gap $\Delta/2$ between  localized impurity levels
and the conduction band.
The energy of polarons on impurity centers is given by
\begin{equation}
E_{i\uparrow(\downarrow)} = -{\Delta\over 2} - (+)\frac{1}{2}V_{ip}m
-(+)\mu_B H,
\label{eq:Ei}
\end{equation}
where $V_{ip}$ is the exchange interaction between localized and
delocalized polarons.
The band diagram for this case is the same as in Fig.~\ref{fig:BndDgr}
with the replacement of the bipolarons by localized polarons.

Assuming that the Hubbard repulsion prevents a double occupancy of the
impurity centers, one can easily obtain the thermodynamic potential
for impurities
\begin{equation}
\Omega_i = -k_BT\ln\left[ 1+2\nu y e^{\Delta/
2k_BT}\cosh\left({\frac{1}{2}V_{ip}m+\mu_BH\over k_BT}\right)\right]
\label{eq:Omi}
\end{equation}
The chemical potential is found to be
\begin{equation}
y = {x-n \over{ 2\nu n \cosh(\zeta)}} \exp(-\delta/t),
\label{eq:yP}
\end{equation}
where $\zeta = (\frac{1}{2}V_{ip}m + \mu_BH)/k_BT$,
if we assume that the total number of impurity states is
$x$($\equiv$ doping).

We have found similar features of the phase transition in zero field
in the impurity case, as compared with the previous case with
bipolarons. Thus, we obtain, by linearizing the system of equations of
state (\ref{eq:n})-(\ref{eq:sig}) with (\ref{eq:yP}), the following
equation for the polaron density at the transition  in zero field:
\begin{equation}
n_c^{1/2} \ln{x-n_c\over{ n_c^2}}=2^{1/2}\delta.
\end{equation}
Apparently, it has solutions only for
$\delta<\delta_{c}(x)$. Therefore, the transition is first order for
$\delta\equiv \Delta/J_{pd}S>\delta_{c}(x)$
and second order for $\delta<\delta_{c}(x)$,
with $\delta_{c}(x)$
slightly larger than in the case of the bipolaron localization.
This follows from the same consideration as in our previous discussion
of Eq.~(\ref{eq:nc}).

The field sensitivity in the case of disorder localized polarons is
much lower than for the bipolarons. This stems from the different
functional dependence of the chemical potential. The present
approximations are valid in the limit $y \ll 1$, meaning that the
polaron carriers are non-degenerate. In contrast to the case of the
bound polaron pair formation, in the impurity case the expression
(\ref{eq:yP}) for $y$ is singular, $y \propto 1/n$, in the limit of
small polaron density. This means that in the vicinity of the
current
carrier density collapse the value of $y$ sharply increases in the
case of polarons localized on impurities. As a result, the collapse
becomes less pronounced, and transport becomes far less sensitive to
an external field.
We note also that Eq.~(\ref{eq:yP}) contains a factor depending on
the external magnetic field in the denominator.  This is in contrast with
the case of bipolarons  (\ref{eq:yt}), where the field
dependence is suppressed by a small factor
$\exp(-J_{st}/k_BT_c)$. This field dependence, however, is small
since always $\mu_B H/k_BT_c \ll 1$, and it quickly vanishes in
the low-temperature phase when the exchange interaction sets in ($m \neq
0$), as one can see from the expression for the parameter $\zeta$ above.
The singular behavior of $y$ as a function of
the density for  $n\rightarrow 0$,
and the Zeeman splitting of the impurity states
makes the transition far less sensitive
to the magnetic field. As a result,
{\em no quantitative description
of the experimental CMR data has been found with the
localization of polarons due to disorder}.

\section{Conclusion}

In conclusion, we have developed a theory
of the ferromagnetic-paramagnetic phase transition in doped
magnetic charge-transfer insulators with
a strong electron-phonon coupling. We have found that  a few
non-degenerate polarons in the $p$ band   polarize
localized $d$ electrons
 because of the huge density of states in the narrow polaronic band. For a
 sufficiently large
 $p-d$ exchange
 $J_{pd}S>\Delta$,  we have obtained
a {\em current carrier density collapse}  at the transition owing to the
formation of immobile local pairs
 in the paramagnetic phase with the binding energy $\Delta$
 about twice that of the polaron level shift\cite{alemot}.
Depending on the ratio $\Delta/(J_{pd}S)$, the transition is
first or second order\cite{ab_cmr1}.

We have explained
the resistivity peak and the colossal
magnetoresistance  of  doped  perovskite manganites,
Fig.~\ref{fig:rte}, as the
result of the current carrier density collapse
due  to the binding of polarons into local pairs (bipolarons).
The density of these immobile pairs has a sharp
peak at the ferromagnetic transition when the system is cooled down through
the critical temperature $T_c$. Below $T_c$ the binding
of polarons into pairs competes with the ferromagnetic exchange
of $p$-holes with the Mn $d^4$ local moments,
which tends to align the polaron moments and, therefore, breaks
those pairs apart.
The spin-polarized polaron band falls below the bipolaron band upon
decrease in temperature, so that all
carriers are unpaired at $T=0$ if $J_{pd}S\geq \Delta$.
Above $T_c$, the bipolaron density decreases because of thermal
activation across the polaron binding energy.
These competing interactions lead to the unusual
behavior of CMR materials,  the huge sensitivity of their transport to
external field, and the very large negative magnetoresistance.

There is a crossover around the transition temperature from polaron
tunneling at low temperatures to polaron hopping, where the latter
dominates at high temperatures.
This explains the  temperature behavior of the resistivity in a wide
temperature range around the transition.
The ferromagnetic to paramagnetic transition is
also accompanied by a sharp anomaly in the specific heat.

The present theory provides a natural explanation for the temperature
dependent gap feature in tunneling spectra\cite{biswas}
and the giant isotope effect
on the temperature of the ferromagnetic transition\cite{mul}.
One of our main conclusions
is that the highly polarized
ferromagnetic phase of manganites is a polaronic doped
semiconductor rather than a  metal.

We expect that the present theory is general enough to also account for
the giant magnetoresistance observed in pyrochlore
manganites\cite{shi} and other systems\cite{ram}.
It is worth mentioning in this regard that the present theory
requires the presence of strong electron-phonon coupling of any origin,
but it  does {\em not} require the presence of Jahn-Teller distortions and/or
the double exchange mechanism. Note that the Jahn-Teller distortions and the
double exchange mechanism are certainly absent in, for instance, pyrochlore
manganites, chromium spinels\cite{ram}, and other CMR systems, so that
the ideas based on the double exchange cannot be applied there at all.
It is believed that at least in perovskite manganites the
local Jahn-Teller distortion may be involved in defining the  crystal
structure of the parent insulating phases\cite{good},
although tilting distortions of MnO$_6$ octahedra are just a result of
steric conditions \cite{geller,ram}. Apparently, the ratio of the sum of
Mn and O ionic radii, $r_{\rm Mn} + r_{\rm O}$, and $(r_{\rm La} + r_{\rm
O})/\sqrt{2}$ (misfit parameter) substantially
differs from unity to make a cubic
structure unstable and favor a rotation of MnO$_6$ octahedra
\cite{van,geller}. 
The tetragonal distortion of MnO$_6$ is large, its symmetry
corresponds to a notion of the Jahn-Teller local distortion.
However, since the steric interaction is strong, it
necessarily deforms the lattice, thus rendering the Jahn-Teller derivation,
strictly speaking, inapplicable. 

It is also believed that
doping by divalent metals introduces holes into the Mn$^{3+}$ $d$-shell,
since the doped systems are less distorted \cite{booth}.
This argument, which may have
supported the relevance of the double exchange mechanism for at least
perovskite manganites,
contradicts the site-sensitive spectroscopic probes \cite{ju,saitoh},
which show unambiguously that holes reside on O sites.
It also neglects two important facts, that (i) the doping is heavy
($\gtrsim 10^{21} e/$cm$^3$) and
there is a substantial size difference between the impurity and host
atoms and (ii) the O $p$-holes
are hybridized with the $d$ states on Mn$^{3+}$,
depending on the value of the charge-transfer gap. Both effects,
together with screened Coulomb hole-hole repulsion, can
apparently explain the observed changes in the lattice distortion
upon doping without invoking the Jahn-Teller mechanism.
These short-range interactions may well be responsible
for the charge-ordered phases observed at some doping levels in
manganites \cite{ram}.
It would be interesting, in this regard, to perform
quantum-chemical calculations of MnO$_6$ clusters with holes doped
onto O site(s). 

Changes and the amount of disorder in the bond lengths are very important
for characterizing the properties of polaronic systems. The reduction in 
bond length distribution width as a result of cooling 
through $T_c$ in doped manganites
has been attributed to (at least partial) delocalization of doped carriers
in low-temperature `metallic' phase. Since the data shows that the carriers
retain their polaronic character below $T_c$, and the residual width of 
the Mn-O bond length
distribution remains larger than that of CaMnO$_3$\cite{booth}, where
the Jahn-Teller Mn$^{3+}$ ions are absent,
the reduction of the width should be mainly
related to instability of bipolarons in this temperature region. Breaking of
polaron bound pairs below $T_c$ may result in a reduction of bond  length
distribution width, and we shall address this question elsewhere.
It is worth repeating that whether or not 
the Jahn-Teller distortions play any role in doped perovskite manganites and 
the exact  location of the carriers is of no importance for the
present scenario  of the CMR.

We acknowledge useful discussions with A.R. Bishop,
D.M. Edwards, J.P.~Franck, K.M. Krishnan,  P.B.~Littlewood, S. von
Molnar, V.G. Orlov, W.E.~Pickett, D.J.~Singh,
S.A.~Trugman, and R.S.~Williams. We especially grateful to
G.~Aeppli, D.S.~Dessau, M.F.~Hundley,  H.-T.~Kim, A.P.~Ramirez,
and G.-m.~Zhao  for useful discussions and  communicating their data.



\begin{references}
\bibitem{twfac}
Otherwise, there will be a prefactor
$\frac{1}{2}d(T/W)^{d/2}$$P(d/2,W/T)\Gamma(d/2)$ in Eq.~(\ref{eq:n})
for $n$, where $d$ is the dimensionality of the system, $\Gamma(x)$ is
the gamma-function and $P(a,x)$
is the incomplete gamma-function. This factor does not change our
conclusions.
\end{references}
\end{document}